\documentclass[LaTeX,twocolumn,epjc3]{svjour3}

\usepackage[linktocpage,breaklinks]{hyperref}
\usepackage{comment}
\usepackage{float}  
\usepackage{newtxmath}
\usepackage{newtxtext}
\usepackage{cuted}
\usepackage{xcolor}  
\definecolor{romared}{RGB}{142,0,28}
\hypersetup{colorlinks=true,
            citecolor=romared,
            linkcolor=romared,
            urlcolor=romared}
\RequirePackage[T1]{fontenc}

\smartqed  
\RequirePackage{graphicx}
      
\RequirePackage{flushend}
\RequirePackage[numbers,sort&compress]{natbib}

\journalname{Eur. Phys. J. C}
\RequirePackage{fix-cm}

\smartqed  
\RequirePackage{graphicx}
\usepackage{physics}
\usepackage{verbatim}
\usepackage[numbers,sort&compress]{natbib}

\usepackage{cuted}

\journalname{Eur. Phys. J. C}
\begin{document}

\title{Relativistic tidal divergences in circular orbits and the dynamics of light rings}

\author{
Victor F. C. Vieira\thanksref{e1,addr1}
        \and
Rafael P. Bernar\thanksref{e2,addr1}
        \and
Caio F. B. Macedo\thanksref{e4,addr1,addr2}
}

\thankstext{e1}{victor.vieira@icen.ufpa.br}
\thankstext{e2}{rbernar@ufpa.br }
\thankstext{e4}{caiomacedo@ufpa.br}

\institute{Programa de Pós-Graduação em Física, Universidade Federal do Pará, 66075-110, Belém, PA, Brazil\label{addr1}
          \and
          Faculdade de Física, Campus Salinópolis, Universidade Federal do Pará, 68721-000, Salinópolis, Pará, Brazil\label{addr2}}
\date{Received: date / Accepted: date}

\maketitle

\begin{abstract}
Tidal forces acting on orbiting bodies arise from inhomogeneities in the gravitational field, generating stresses that can deform or even disrupt these objects. In this work, we analyze relativistic tidal forces associated with ultracompact objects described by static and spherically symmetric spacetimes, focusing on observers in circular geodesic motion. We show that, in contrast to the case of radial geodesics, tidal forces diverge as the orbit approaches null circular geodesics. As illustrative examples, we study two uniform-density stellar models: one isotropic and another supported purely by tangential stresses. We conjecture that the divergence of tidal forces near light rings may play a role in the nonlinear stability of ultracompact, horizonless objects.

\end{abstract}

\section{Introduction}
While black holes (BHs) became a paradigm in physics, mostly through their role in explaining the supermassive objects at the center of the galaxies, some argue that it is impossible to directly detect the event horizon (see, e.g., Ref. \cite{Testing_Cardoso_2019} for a review).  The observational evidence for these objects comes from the dynamics of the orbits around them, the effects caused by the presence of shadows, and the gravitational waves that are emitted \cite{2019,PhysRevLett.116.061102}. However, many of the features originally credited to BHs can be mimicked by so-called ultracompact objects (UCO): astrophysical objects with radii close enough to the would-be horizon radius \cite{marks2025longtermstablenonlinearevolutions,PhysRevLett.119.251102,pani2009ergoregioninstabilityblackhole}. Therefore, UCOs can act as alternatives to BHs, without having event horizons and the many problems they possibly carry.

The possible existence of ultracompact objects without a horizon motivates the study of questions such as the stability of these structures \cite{cunha2017light}. Among the astrophysical features relevant in this context, a significant one is the presence of light rings, which are circular orbits of photons around the object \cite{PhysRevD.105.064026}. Note that these configurations can admit the existence of two light rings: an unstable\footnote{Note that the word ``unstable'' here refers to the fact that the circular orbits generating the light ring are unstable.} one outside the object and a stable ring of light inside the star (a characteristic absent in black holes) \cite{guo2024nonlinear}. Light rings are crucial for understanding the image formation around compact objects, such as the ones coming from the event horizon telescope~\cite{Rosa:2023qcv,Rosa:2024bqv}.

Light rings not only affect the radiation around ultracompact objects, but are also orbital motion that, depending on their stability, can influence the evolution of physical structures in their vicinity \cite{cunha2025backreaction,Di_Filippo_2024}. In regions near the stable circular orbits of photons, the existence of long-lived perturbations might influence the system's stability \cite{PhysRevD.90.044069}, akin to the turbulent instability of anti-de Sitter space~\cite{Bizon:2011gg}. For the case of ultracompact objects, this potential instability depends on the symmetry of the perturbations, which is connected to the light rings in the eikonal limit. In order for the collapse to occur, however, one must provide conditions for the formation of such clumps of matter~\cite{PhysRevD.90.044069}, meaning the existence of orbital motions within the star that could support such transient states.

The stability of fluid objects in orbital motion is linked to the tidal forces/deformations generated by the gravitational field. It is natural, therefore, to analyze the tidal forces that ultracompact objects generate in their interior in view of possible implications for the accumulation of matter in the interior of the star. 

In the relativistic framework, which is our primary focus, tidal forces can be computed using the geodesic deviation equation, projected onto a specific local frame of interest. Physically, these forces arise from the inhomogeneity of the gravitational field experienced by a body and become particularly significant near massive objects such as neutron stars and black holes \cite{Lima_Junior_2022}. In the relativistic regime, tidal forces can be substantially amplified, especially as the orbit approaches null circular geodesics.

When a stellar object orbits a black hole, it is important to consider how tidal forces act on this object and how its structures and internal physical quantities, such as pressure and density, respond to these forces. The studies described in Refs.\cite{chandrasekhar1969ellipsoidal,stockinger2024relativisticrocheproblemstars,Wiggins_2000,1973ApJ...185...43F} analyze how the physical structure of the object in circular orbit responds to tidal effects. In particular, Ref. \cite{stockinger2024relativisticrocheproblemstars} discusses the conditions under which the orbiting object manages to maintain its structural integrity, establishing a critical value beyond which the object's cohesion is no longer sustained, called the Roche limit.

In this work, we examine the tidal forces associated with circular orbits around ultracompact objects. Outside the star, since we consider spherically symmetric configurations, the results coincide with those of the standard Schwarzschild spacetime. Inside the star, there exist timelike circular orbits that, in the high-energy and high-angular-momentum limit, are connected to the light rings. We investigate the tidal forces associated with these orbital motions inside the star, motivated by the possibility that they might influence the trajectories of localized matter clumps that form within the interior. Notably, the tidal forces always diverge at both the inner and outer light rings, suggesting that gravitational collapse may be forbidden in such configurations.

We illustrate the cases of ultracompact objects based on fluid models \cite{PhysRevD.99.104072}, focusing on static and spherically symmetric configurations \cite{chandrasekhar1998mathematical,carroll2019spacetime,hobson2006general,d2023introducing,PhysRev.55.364}. At first, we leave all the computations in terms of pressure and density matter distribution, showing that the tidal force divergences are generic. We then use two simple models describing isotropic and anisotropic uniform density fluid stars \cite{BeL,florides1974new}. While isotropic stars are restricted by the Buchdahl limit $R/M>9/4$ \cite{PhysRev.116.1027,Sharma_2021}, the Florides solution allows for more compact configurations.

In this scenario, we use the covariant formalism of general relativity to describe the behavior of these forces from a geometric point of view, in a curved space-time. We use the mechanism of geodesic deviation, which describes the variation between the distance of two bodies as they evolve during geodesic motion \cite{Fuchs90,Madan:2022spd}. Thus, we aim to understand how the relative acceleration between parts of an extended body behaves under the gravitational influence of an ultracompact object, verifying the implications of these effects on the structure of the body and how this is associated with the stability of the object.

The remainder of the paper is organized as follows. In Sec.~\ref{sec:geodesic} we review circular geodesic motion in a spherically symmetric spacetime, mostly on time-like geodesics, but also discuss light rings and how to find them. In Sec.~\ref{sec:deviation} we discuss the geodesic deviation equation in spherically symmetric spacetimes, projecting the equations onto circular orbiting frames, showing the divergences in the light-ring limit. In Sec.~\ref{sec:models} we apply the results to two different cases, namely iso\-tro\-pic and anisotropic uniform density stars, analyzing the intensity of the tidal forces for these. Finally, in Sec.~\ref{sec:conclusion} we present our main conclusions and perspectives. In the remainder of this work we use natural units such that $c=G=1$ and signature $(-,+,+,+)$.

\section{Geodesics in static spherically symmetric spacetimes}\label{sec:geodesic}

In this work, we will focus on spherically symmetric and static configurations for ultracompact objects, which are described by the metric
\begin{equation}\label{eq:metric}
    ds^2 = -f(r)dt^2 + h(r)dr^2 + r^2d\Omega^2,
\end{equation}
where $f(r)$ and $h(r)$ are functions of the radial coordinate only and $d\Omega^2= d\theta^2 + \sin^2\theta d\varphi^2$ is the line element of the unit 2-sphere. We shall also demand that $f(r)$ and $h(r)$ be such that Einstein's equations are satisfied, but for the moment we shall treat them generically. 

 Geodesics on the equatorial plane of this spacetime can be obtained through the following Lagrangian:  
\begin{equation}\label{6}
    \mathcal{L} =\frac{1}{2}\left[- f(r) \dot{t}^2 + \frac{1}{h(r)} \dot{r}^2 + r^2 \dot{\varphi}^2 \right]\ ,
\end{equation}
 where the dot denotes derivative with respect to the affine parameter. As the Lagrangian is independent of both $t$ and $\varphi$, this implies that the momenta of the two coordinates, i.e. the specific energy $E=-\partial\mathcal{L}/\partial\dot{t}$ and angular momentum $L=\partial\mathcal{L}/\partial\dot{\varphi}$, are conserved along the trajectory. Hence, 
 \begin{equation}\label{7}
     \dot{\varphi}=\frac{L}{r^2}, ~~ \dot{t}=\frac{E}{f(r)}.
 \end{equation} 
Thus, for geodesic orbits in the equatorial plane, we have to solve
\begin{align}\label{eq:radial}
     \frac{f(r)}{h(r)}\dot{r}^2=L^2\left[\frac{E^2}{L^2}-V_{eff}(r)\right],
\end{align}
where the  effective potential $V_{eff}(r)$ is given by
\begin{align}\label{eq: Vmr}
 V_{eff}(r)=f(r)\left(\frac{L^2}{r^2}+\epsilon\right),
\end{align}
with $\epsilon=0\ (1)$ describing null (time-like) geodesics.

We now turn our attention to circular geodesics, which are particularly interesting in view of tidal forces for objects in circular motion \cite{Fuchs90}. Considering time-like geodesics, $\epsilon=1$, and that $\dot{r}=0$ and $\ddot{r}=0$, we find from Eq.~\eqref{eq:radial} and its derivative that the energy and angular momentum for circular orbits with radius $r_c$ are given by
 \begin{equation}\label{eq:e_l_timelike}
E_c= \frac{\sqrt{2} f(r_c)}{\sqrt{2 f(r_c)-r_c f^\prime(r_c)}},\  L_c= \frac{r_c^{3/2} \sqrt{f^\prime(r_c)}}{\sqrt{2 f(r_c)-r_c f'(r_c)}}.
 \end{equation}
Circular time-like geodesics exist provided that the specific energy and angular momentum given by Eq. \eqref{eq:e_l_timelike} are real. Note, however, that this does not explicitly state anything about the \textit{stability} of such circular orbits. We can analyze the stability through the effective potential. If $V_{eff}^{\prime\prime}(r_c)>0$ the orbit is stable, $V_{eff}^{\prime\prime}(r_c)<0$, the orbit is unstable and $V_{eff}^{\prime\prime}(r_c)=0$ it is marginally stable. For instance, considering the Schwarzschild exterior spacetime, one finds that we have stable orbits for $r_c>6M$, unstable orbits in the range $3M<r_c<6M$, and a marginally stable orbit at the transition $r_c=6M$, which is also called the innermost stable circular orbit (ISCO) for the case of the Schwarzschild BH. Although ultracompact stars have the same Schwarzschild exterior and, as such, the same orbital analysis outside the star, there are possible \textit{stable} circular orbits within the star. We shall further discuss these orbits later on.

We note that the expressions for the specific energy and angular momentum for circular orbits, given by Eq.~\eqref{eq:e_l_timelike}, formally diverge when $2f(r_c)-r_cf^{\prime}(r_c)\to0$. We shall see in the next section that this is linked to a particle approaching the null geodesic limit, that is, the case of ultra-relativistic particles.

\section{Light rings}
Light rings are null orbits around the ultracompact object that express extreme deflection of the light rays. As mentioned above, since ultracompact objects do not have an event horizon, the existence of geodesic orbits inside the star is also possible, as these stars are compact enough to support them. 

We can analyze the existence of light rings directly through the effective potential.  Considering the case $\epsilon=0$, we have that $V_{eff}=L^2f(r)/r^2$. Note that the $L^2$ in the effective potential only amounts for a scaling factor, and does not change the position of extrema. The existence of local maxima in the potential correspond to the presence of unstable light rings \cite{Ghosh_2021}. Alternatively, local minima indicate the presence of a stable light rings. Considering the potential for null geodesics, we have that the extrema location $r_l$ are given by
\begin{equation}\label{eq:loclr}
    2f(r_l)-r_lf^{\prime}(r_l)=0.
\end{equation}
This condition is used to determine the positions of light rings in spherically symmetrical systems. As we predicted, this is precisely the location occuring the divergence of the specific energy and angular momentum of massive particles in circular motion. 

For ultracompact objects, it is well know that light rings come in pairs of stable and unstable light rings \cite{cunha2017light}. For the cases we shall explore here, there is only one pair. The outer light-ring (unstable), which we shall denote by $r_+$, is the same as in Schwarzschild spacetime, as we have that the exterior is vaccuum and the radius of the star is $R<3M$. Therefore, we have $r_+=3M$. The inner light-ring (stable) is \textit{within} the star, and depends on the particular model we are dealing.

We can verify the stability of the light rings by analyzing, once again, the second derivative of the effective potential. Outside the star, the solution of equation~\eqref{eq:loclr} provides the unstable light ring, since $V_{eff}^{\prime\prime}(r_+)<0$. This can easily be verified by using the Schwarzschild metric. On the other hand, we shall see that for the light ring located inside the ultracompact object, we have $V_{eff}^{\prime\prime}(r_-)>0$, implying a stable orbit \cite{PhysRevD.90.044069}. Therefore, simply from a plot of the effective potential one can verify the existence of light rings: minima are stable ones and maxima are unstable ones.

\begin{figure}
    \centering
    \includegraphics[width=0.8\linewidth]{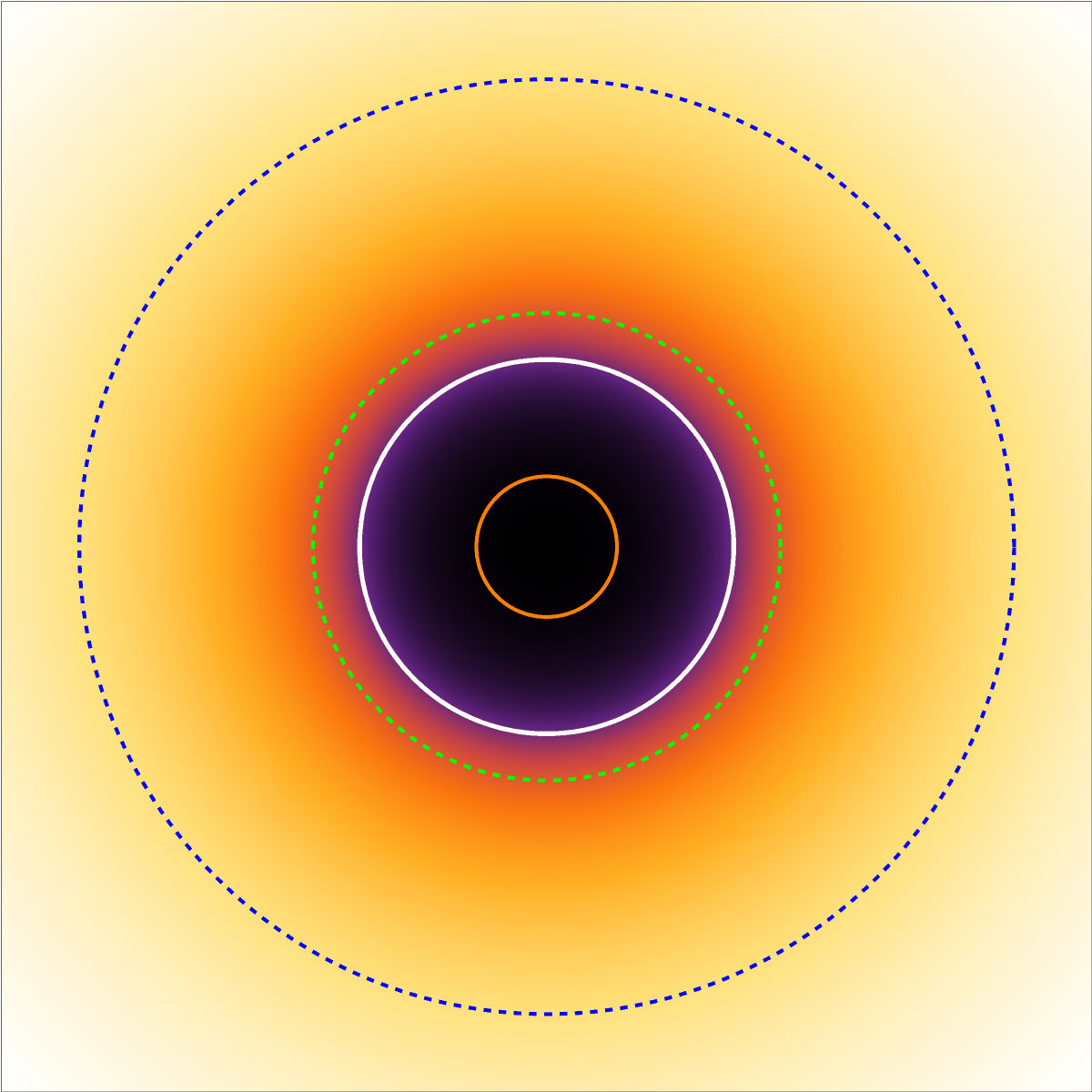}
    \caption{Illustrative picture of ultacompact objects. The marginally stable circular orbit, at $r = 6M$, is indicated by the dashed blue circle. The outer unstable light ring, at $r = 3M$, appears in green. For objects without a horizon, there may be an inner stable light ring, represented in orange, whose position depends on the star's radius. In this example, we have an uniform density isotropic star of radius $R = 2.4M$.}
    \label{fig:figureOUC}
\end{figure}

We illustrate the ``topography'' of the stellar models we are dealing with in Fig.~\ref{fig:figureOUC}. The surface of the star is represented by the white curve. We can visualize the arrangement of the light rings, the stable one  represented by the orange curve, and the unstable light ring by the dashed green curve. The marginally stable circular orbit is represented by the blue dashed line. In addition, the color gradient in the figure illustrate the gravitational \textit{redshift} associated with the ultracompact object's field.

\section{Geodesic deviation and tidal forces in circular motion}\label{sec:deviation}

The tidal forces acting on an object can be interpreted as a consequence of the inhomogeneity of the gravitational field \cite{Hong_2020}. This variation in the intensity of the gravitational field generates a differential force, resulting in internal stresses within the object, leading to deformations \cite{stockinger2024relativisticrocheproblemstars}. In extreme cases, such as objects near black holes, these forces can become strong enough to fragment the object.

In the context of general relativity, we describe tidal forces through a mechanism called geodesic deviation (see, e.g., \cite{hobson2006general,thorne2000gravitation} for textbook materials on the subject). Assuming two nearby particles with their respective geodesic trajectories, we can consider the existence of a separation vector, $\xi^\mu$, between them. The geodesic deviation equation describes by the rate of change of this vector over the particle's proper time $\tau$, being given by
\begin{equation}\label{eq:dGeo}
    \frac{D^2\xi^\mu}{D\tau^2} = R^\mu_{\nu\sigma\rho}u^\sigma u^\nu \xi^\rho,
\end{equation}
where  $u^\sigma$\ is the velocity vector of the particle. 

The second-order derivative of $\xi^\mu$ with respect to the proper time describes the rate of change of the speed of separation between the particles, i.e. it is an acceleration. We can check how this acceleration acts on this extended body. To do this, we take a local reference point on one of the particles and introduce a tetrad formalism \cite{mitsou2020tetrad,thorne2000gravitation,chandrasekhar1998mathematical}. Thus, for each point in space-time, we associate a set of orthogonal basis vectors \cite{zakharovTetradFormalismReference2006,Wulandari:2024gia}, associated to a specific moving frame. Using this mechanism, we can decompose the components of the 4-velocity in the local reference frame, given by $u^{\hat{\alpha}}$, where we represent the projected quantities by a hat over the index. Therefore, taking the particle's local reference point, let us choose a coordinate system so that it is in a circular orbit. For this purpose, we have that the 4-velocity given by of a particle in circular geodesic is given by
\begin{equation}
u^{\mu}=\left(\dot{t},0,0,\dot{\varphi}\right)=\left(\frac{E_c}{f(r_c)},0,0,\frac{L_c}{r_c^2}\right).
\end{equation}
We shall use the four-velocity as the first vector and choose the remaining such that it obeys $\eta^{\hat{a}\hat{b}}=\hat{e}^{\mu\hat{a}}{\hat{e}_\mu}^{~~\hat{b}}$. We have
\begin{equation}
    \hat{e}_{\hat{t}}^\mu=\alpha^{-1/2}\left(1,0,0,\sqrt{\frac{f-\alpha}{r^2}}\right),
\end{equation}
\begin{equation}
    \hat{e}_{\hat{r}}^\mu=\left(0,\sqrt{h},0,0\right),
\end{equation}
\begin{equation}
    \hat{e}_{\hat{\theta}}^\mu=r^{-1}\left(0,0,1,0\right),
\end{equation}
\begin{equation}
    \hat{e}_{\hat{\varphi}}^\mu=\alpha^{-1/2}\left(\sqrt{1-\frac{\alpha}{f}},0,0,\frac{\sqrt{f}}{r}\right).
\end{equation}\label{19}

\noindent where we have defined $\alpha=f-\frac{1}{2} r f^\prime$. Notice that $\alpha\to0$ at the light-ring position. 
In circular orbits, the tetrad base above is carried along the particle's trajectory, so it is not a constant \cite{1983RSPSA.385..431M}. This means that the left-hand side of the \eqref{eq:dGeo} equation will have additional terms related to Coriolis and centrifugal acceleration \cite{1990AN....311..271F}. In this way, we can see that the components of the relative acceleration of the separation vector will take the basic form given by
\begin{align}
\ddot{\xi}^{\hat{i}}=P^i{}_j\dot{\xi}^{\hat{j}}+K^i{}_{j}\xi^{\hat{j}},
\end{align}
where $P^i{}_j$ are the coefficients proportional to the first time derivative of the separation vector, associated with the inertial forces. While $K^i{}_j$ describe the coefficients proportional to the separation vector between the particles, associated with the acting tidal effects. Using the components for the curvature tensor, we can substitute them into the geodesic deviation equation \eqref{eq:dGeo} \cite{Philipp_2019}.

For the case of circular orbits, the tetrad base is not parallelized and, therefore, it is necessary to use a modified geodesic deviation equation, which is written in terms of \textit{spin-connections} (see, e.g., Ref. \cite{carroll2019spacetime} for a introductory account on the subject). This property indicates that there is a local rotation of the base along the trajectory, which directly affects how we calculate covariant derivatives of vectors in a way that is compatible with the observer's local referential \cite{1973ApJ...185...43F}. Thus, using the geodesic deviation equation expressed in the form \eqref{eq:dGeo} is not sufficient to correctly capture the tidal effects perceived in the tetradic reference frame. It is necessary to rewrite this equation in terms of the local orthonormal basis. However, due to the non-conservation of the chosen basis, we must adapt the covariant derivative, replacing the Christoffel connection $\Gamma^\alpha_{\mu\nu}$ with the spin connection $\omega_\mu{}^{\hat{a}}{}_{\hat{b}}$, given by
\begin{align}
    \omega_\mu{}^{\hat{a}}{}_{\hat{b}} = e^{\hat{a}}{}_\lambda \Gamma^\lambda_{\mu\nu} e^\nu{}_{\hat{b}} - e^{\hat{a}}{}_\lambda \partial_\mu e^\lambda{}_{\hat{b}},
\end{align}
 which incorporates the local variation of the tetrads \cite{de_Andrade_2001}. 
 
 The spin-connection formalism allows us to adopt any orthonormal basis, even if its components are not parallelized. In this way, we adopt the format of the covariant derivative expressed in terms of the spin-connection. Thus, by applying this modification to the left-hand side of the geodesic deviation equation \eqref{eq:dGeo}, we were able to include the components of the base in the derivative, i.e.
\begin{align}
\nabla_\mu\nabla_\nu\left(\xi^{\hat{b}} e_{\hat{b}}\right)=e_{\hat{b}}\left[\partial_\mu \partial_\nu \xi^{\hat{b}}+\partial_\mu \xi^{\hat{c}} \ \omega_\nu{}^{\hat{b}}{}_{\hat{c}} +\partial_\nu \xi^{\hat{c}} \ \omega_\mu{}^{\hat{b}}{}_{\hat{c}}
    \nonumber\right.\\-\left.\Gamma^\lambda_{\mu\nu} \ \partial_\lambda \xi^{\hat{b}} + \ \xi^{\hat{c}}\left(\partial_\mu \omega_\nu{}^{\hat{b}}{}_{\hat{c}} -\Gamma^\lambda_{\mu\nu} \ \omega_\lambda{}^{\hat{b}}{}_{\hat{c}}+\omega_\nu{}^{\hat{d}}{}_{\hat{c}} \ \omega_\mu{}^{\hat{b}}{}_{\hat{d}} \right)\right].
\end{align}

Using the significant components of the curvature tensor, we can substitute them into Eq.(\ref{eq:dGeo}). Thus, we find the following equations for the tidal forces on references in circular orbits:
\begin{align}
    \dv[2]{\xi^{\hat{r}}}{\tau}-&\sqrt{\frac{2 h f^{\prime}}{r f}}\dv{\xi^{\hat{\varphi}}}{\tau}-\left[\frac{h f^{\prime}}{2 f r}+\frac {h{f^{\prime}}^2 - 2 fhf^{\prime\prime}} {4f^2-2rff^\prime}\right]\xi^{\hat{r}}=0,\label{eq:tfmetric1}\\
    \dv[2]{\xi^{\hat{\theta}}}{\tau}-&\left[\frac{f^{\prime}(r)}{2 r f(r)-r^2 f^{\prime}(r)}\right]\xi^{\hat{\theta}}=0,\label{eq:tfmetric2}\\
    \dv[2]{\xi^{\hat{\varphi}}}{\tau}+&\sqrt{\frac{2 h f^{\prime}}{r f}}\dv{\xi^{\hat{r}}}{\tau}=0.\label{eq:tfmetric3}
\end{align}

From the components of the tidal forces for circular orbits, it is possible to observe their behavior in orbits located close to the light rings. Substituting the Schwarzschild metric, we can see this behavior in the unstable light ring, located outside the star, using the equations
\begin{align}\label{eq:tce1}
    \dv[2]{\xi^{\hat{r}}}{\tau}=&2\sqrt{\frac{M}{r^3}}\dv{\xi^{\hat{\varphi}}}{\tau}+\left[\frac{3 M (2 M-r)}{r^3 (3 M-r)}\right]\xi^{\hat{r}},\\\label{eq:tce2}
    \dv[2]{\xi^{\hat{\theta}}}{\tau}=&\left[\frac{M}{r^2 (3 M-r)}\right]\xi^{\hat{\theta}},\\
    \dv[2]{\xi^{\hat{\varphi}}}{\tau}=&-2\sqrt{\frac{M}{r^3}}\dv{\xi^{\hat{r}}}{\tau}.\label{eq:tce3}
\end{align}
The terms proportional to the first derivative are connected to inertial forces felt by the moving body. The tidal force -- proportional to the displacement vector -- formally diverge at $r=r_+=3M$, i.e., at the position of the light-ring. However, in the external spacetime, this region can be considered unphysical as the orbital motion in that region are not stable.

For the orbital motion inside the star, we need to specify additional requirements for the metric coefficients $(f(r),h(r))$. In the following, we shall impose that they are described by the Einstein's equation with an anisotropic fluid matter.

\section{Models of  ultracompact fluids stars}\label{sec:models}

The matter that composes the ultracompact stellar object will be described on the basis of an anisotropic fluid model \cite{BeL,PhysRev.55.364}. That is, we consider that the radial and the tangential pressure are different. In this way, we have
\begin{equation}
    T_{\mu\nu} = \rho v_\mu v_\nu + p k_\mu k_\nu + p_\bot \Pi_{\mu\nu},
\end{equation}
where $v_\mu$ is the 4-velocity of the fluid element, $\rho$ is the energy density, $p$ is the radial pressure, $p_\bot$ is the tangential pressure, $k_\mu$ is a unit radial vector satisfying $k_\mu k^\mu=1$, and $k_\mu v^\mu=0$ and $\Pi_{\mu\nu}$ is a projection operator, defined by
\begin{equation}
    \Pi_{\mu\nu} = g_{\mu\nu} + v_{\mu} v_{\nu} - k_{\mu} k_{\nu}.
\end{equation}

Considering an anisotropic fluid, we can find the equations that describe a static distribution of a spherically symmetric fluid, which correspond to the generalized \textit{Tolman–Oppenheimer–Volkoff} (TOV) equations \cite{BeL,PhysRev.55.364}, given by
\begin{align}
&p^{\prime}=-(\rho+p)\frac{f^\prime(r)}{2 f(r)}-\frac{2}{r}(p - p_\bot), \label{eq:eqdpdr}\\
&\frac{f^\prime(r)}{f(r)}=\frac{2m+8\pi r^3 p}{r\left(~r-2m\right)}, \label{eq:f'f}\\
  &m^{\prime}=4\pi r^2\rho,\label{eq:dmdr}
\end{align}
where $\prime$ represents radial derivatives and the mass function $m(r)$ is defined via $h(r)=1-2m(r)/r$.

For convenience, we will focus on fluid stars with uniform density. Therefore, we will have that $\rho=\rho_0$ is constant inside the star. We can integrate \eqref{eq:dmdr} with respect to the radial coordinate and obtain
\begin{align}\label{eq:m(r)}
    m(r)=\
    \begin{cases}
        \frac{4}{3}\pi r^3 \rho_0, & r < R, \\
        \frac{4}{3}\pi R^3 \rho_0=M, & r\geq R, 
    \end{cases}
\end{align}
where $M$ is the total (ADM) mass of the ultracompact object.

To better visualize the difference between the tidal forces inside and outside the star, it is convenient to use the field equations to eliminate the metric derivatives in terms of the fluid quantities. Using the equations \eqref{eq:eqdpdr}--\eqref{eq:dmdr}, we get 
\begin{strip}
\begin{align}
    &\dv[2]{\xi^{\hat{r}}}{\tau}=2 \sqrt{\frac{m}{r^3}+4 \pi  p}
 \dv{\xi^{\hat{\varphi}}}{\tau}+
 \left[\frac{r m \left(12 \pi  r^2 p-4 \pi  r^2 (\rho_0 -4 \sigma )-3\right)+6 m^2+4 \pi  r^4 \left(4 \pi  r^2 p (3 p+\rho_0 )+\rho -2 \sigma \right)}{r^3 \left(3 m+4 \pi  r^3 p-r\right)}
    \right]\xi^{\hat{r}},\label{eq:tfcamp1}\\
    &\dv[2]{\xi^{\hat{\theta}}}{\tau}=-\left[\frac{m+4 \pi  r^3 p}{r^2 \left(-3 m-4 \pi  r^3 p+r\right)}\right]\xi^{\hat{\theta}},\label{eq:tfcamp2}\\
    &\dv[2]{\xi^{\hat{\varphi}}}{\tau}=2 \sqrt{\frac{m}{r^3}+4 \pi  p} \ \dv{\xi^{\hat{r}}}{\tau}\label{eq:tfcamp3},
\end{align}
\end{strip}
where we have defined $\sigma=p-p_\bot$ From the above, we directly see the dependence on the matter quantities: vaccuum is obtained by making $\rho=p=\sigma=0$. Clearly, that points to the divergence of the tidal forces at the Schwarzschild light ring position $r_+=3m=3M$. In the following, we specify two other models with divergences on stable light rings within the star.

\subsection{Isotropic configurations}

One of the simplest model that explores the properties of ultracompact objects is the constant-density star. Inside the star, the metric functions are obtained from the field equations \eqref{eq:eqdpdr}--\eqref{eq:dmdr}, we obtain given by
\begin{align}
        f(r)&=\frac{1}{4R^3}\left(\sqrt{R^3-2Mr^2}-3R\sqrt{R-2M}\right)^2 \label{eq:f(r)iso}\\
    h(r) &=\left(1-\frac{2Mr^2}{R^3}\right)^{-1},\label{eq:h(r)iso}\\
     p(r)&=\rho_0 \left(\frac{R\sqrt{R-2M}-\sqrt{R^3-2Mr^2}}{\sqrt{R^3-2Mr^2}-3R\sqrt{R-2M}}\right),
\end{align}
where $R$ is the radius of the star. Outside of the star, the spacetime is described by the Schwarzschild metric. The radius is found using the condition $p(R)=0$.
The central pressure on the object, $p_c=p(r=0)$, will be given by
\begin{align}\label{eq:rhoc}
    p_c=\rho_0\left[\frac{1-\left(1-2M/R\right)^\frac{1}{2}}{3\left(1-2M/R\right)^\frac{1}{2}-1}\right].
\end{align}
 
A particular aspect of isotropic stars is the existence of an upper limit for the ratio between the star's mass and radius, $M/R$, known as the Buchdahl limit.
The Buchdahl limit comes from the fact that central pressure needed to support the structure of the object must remain finite. From Eq. \eqref{eq:rhoc} we obtain that the central pressure blows-up when $R=9M/4$. Therefore, isotropic stars can not approach the Schwarzschild limit through a sequence of equilibrium configurations. Furthermore, given that the Buchdahl limit is less than $3M$, we can consider the existence of ultracompact isotropic objects with $R<3M$.

To analyze the behavior of the tidal forces in this configuration, we now specialize the general expressions of the geodesic deviation equations for the case of a constant density star. This is done by substituting the corresponding metric functions $f(r)$ and $h(r)$, given by equations (\ref{eq:f(r)iso}) and (\ref{eq:h(r)iso}), into the expressions of the tidal components, expressed in (\ref{eq:tfmetric1}-\ref{eq:tfmetric3}).  Otherwise, since the pressure $p(r)$ and the energy density $\rho_0$ are known analytically in this model, we can use these physical quantities for the isotropic case in the equations (\ref{eq:tfcamp1}-\ref{eq:tfcamp3}).  This allows us to explicitly calculate the radial and angular components of the tidal acceleration inside the star. For this isotropic configuration, the tidal forces become
\begin{strip}
\begin{align}
   &\dv[2]{\xi^{\hat{r}}}{\tau}=\left[2 \sqrt{\frac{M \sqrt{1-\frac{2 M r^2}{R^3}}}{R^3 \left(\frac{3}{2} \sqrt{1-\frac{2 M}{R}}-\frac{1}{2} \sqrt{1-\frac{2 M r^2}{R^3}}\right)}}\right]\dv{\xi^{\hat{\varphi}}}{\tau}-\left[\frac{12 M^2 r^2\sqrt{1 - \frac{2 M}{R}}}{R^6\left(3\sqrt{1 -
\frac{2 M}{R}} - \sqrt{1 - \frac{2 M r^2}{R^3}} \right)\left(3
\sqrt{1 - \frac{2 M}{R}}\sqrt{1 - \frac{2 M r^2}{R^3}} - 
     1 \right)}\right]\xi^{\hat{r}},\\
    &\dv[2]{\xi^{\hat{\theta}}}{\tau}=\left[-\frac{2 M}{R^3 \left(3 \sqrt{1-\frac{2 M}{R}} \sqrt{1-\frac{2 M r^2}{R^3}}-1\right)}\right]\xi^{\hat{\theta}},\\
   &\dv[2]{\xi^{\hat{\varphi}}}{\tau}=-\left[2 \sqrt{\frac{M \sqrt{1-\frac{2 M r^2}{R^3}}}{R^3 \left(\frac{3}{2} \sqrt{1-\frac{2 M}{R}}-\frac{1}{2} \sqrt{1-\frac{2 M r^2}{R^3}}\right)}}\right]\dv{\xi^{\hat{r}}}{\tau}.
\end{align}
\end{strip}
From the above equations, we can see the existence of an additional divergence within the star. By using the light ring condition, we can find this position \textit{analytically}, as 
\begin{equation}
    r_- = \frac{R^{3/2} \sqrt{9 M - 4 R}}{\sqrt{18 M^2 - 9M R}}.
\end{equation}

\subsection{Anisotropic configurations and the Florides solution}
The anisotropic approach for stars allows us to study solutions where the radial and tangential pressures are different. However, one still need an additional equation of state to solve the equations. A relation proposed by Bowers and Liang \cite{BeL} $\sigma$ is
\begin{equation}\label{eq:sigma}
    \sigma=\frac{1}{3}\lambda\left(\rho+3p\right)\left(\rho+p\right)\left(1-\frac{2M}{r}\right)^{-1}.
\end{equation}
With the above, an analytical uniform density solution is possible.

By pluging \eqref{eq:sigma} into the differential equations and solving Eq. \eqref{eq:eqdpdr}, we obtain
\begin{align}
    \dv{p}{r}=-\left(\frac{4\pi}{3}-\frac{2\lambda}{3}\right)\left(\rho_0+3p\right)\left(\rho+p\right)\left(1-\frac{8\pi \rho_0 r^2}{3}\right)^{-1}.
\end{align}
Thus, the radial pressure for an anisotropic fluid with constant density is given by
\begin{equation}\label{eq:prq}
    p(r)=\rho_0 \left[ \frac{(1 - 2m/r)^Q - (1 - 2M/R)^Q}{3(1 - 2M/R)^Q - (1 - 2m/r)^Q} \right],
\end{equation}
where $Q=\frac{1}{2}+\frac{\lambda}{4\pi}$ and we define, based on the radius $R$ of the star that, $p(R)=0$, and $m(R)= M$ being the total mass of the star.

In particular, for simplicity, we will consider the case where $\lambda=-2\pi$. This implies that the radial pressure identically zero. Thus, the tangential pressure takes the form 
\begin{equation}
    p_\bot(r)=\frac{2\pi \rho_0^2}{3}\left(1-\frac{8\pi \rho_0 r^2}{3}\right)^{-1}r^2.
\end{equation}
This solution, obtained by Florides, describes an anisotropic fluid in hydrostatic equilibrium,  in which the object is supported exclusively by the tangential contribution of pressure. Outside the distribution of matter, space-time is again described by the Schwarzschild solution. From the remaining field equation, we have that the metric function $f(r)$ will be expressed by 
\begin{equation}
   f(r)=\frac{\left(1-\frac{2 M}{R}\right)^{3/2}}{\sqrt{1-\frac{2 r^2 M}{R^3}}},
\end{equation}
while the function $h(r)$ is the same as that described in (\ref{eq:h(r)iso}).

We can see that for this stellar model supported solely by the tangencial pressure, the Buchdahl limit can be violated, allowing for more compact structures. There is no singular point for a given $R$ at the center of the star, as both $f(r)$ and $\sigma(r)$ are regular there, so we can conclude that we can push the star's radius up to the Schwarzschild limit $R\to2M$.

In the same way, we can obtain the sea forces for the interior solution of anisotropic stars, based on the Florides model. Thus, we obtain
\begin{align}
    \dv[2]{\xi^{\hat{r}}}{\tau}=&2 \sqrt{\frac{M}{R^3}}\dv{\xi^{\hat{r}}}{\tau}-\left[\frac{6 M^2 r^2}{R^6-3 M r^2 R^3}\right]\xi^{\hat{r}},\\
    \dv[2]{\xi^{\hat{\theta}}}{\tau}=&\left[-\frac{M}{R^3-3 M r^2}\right]\xi^{\hat{\theta}},\\\label{eq:tcflo3}
    \dv[2]{\xi^{\hat{\varphi}}}{\tau}=&-2 \sqrt{\frac{M}{R^3}} \ \dv{\xi^{\hat{r}}}{\tau}.
\end{align}
Note that the expressions are quite simple. Similarly to the isotropic case, the tidal forces diverge at the inner light-ring. For the anisotropic configuration described by Florides' solution, location of the stable light ring is given by
\begin{equation}\label{eq:lrflorides}
r_- = \frac{R^{3/2}}{\sqrt{3M}},
\end{equation}
while the position of the light ring outside the object is the same in both cases.

\subsection{Tidal forces for circular orbits in ultracompact stars}

\begin{figure}
    \centering
\includegraphics[width=1\linewidth]{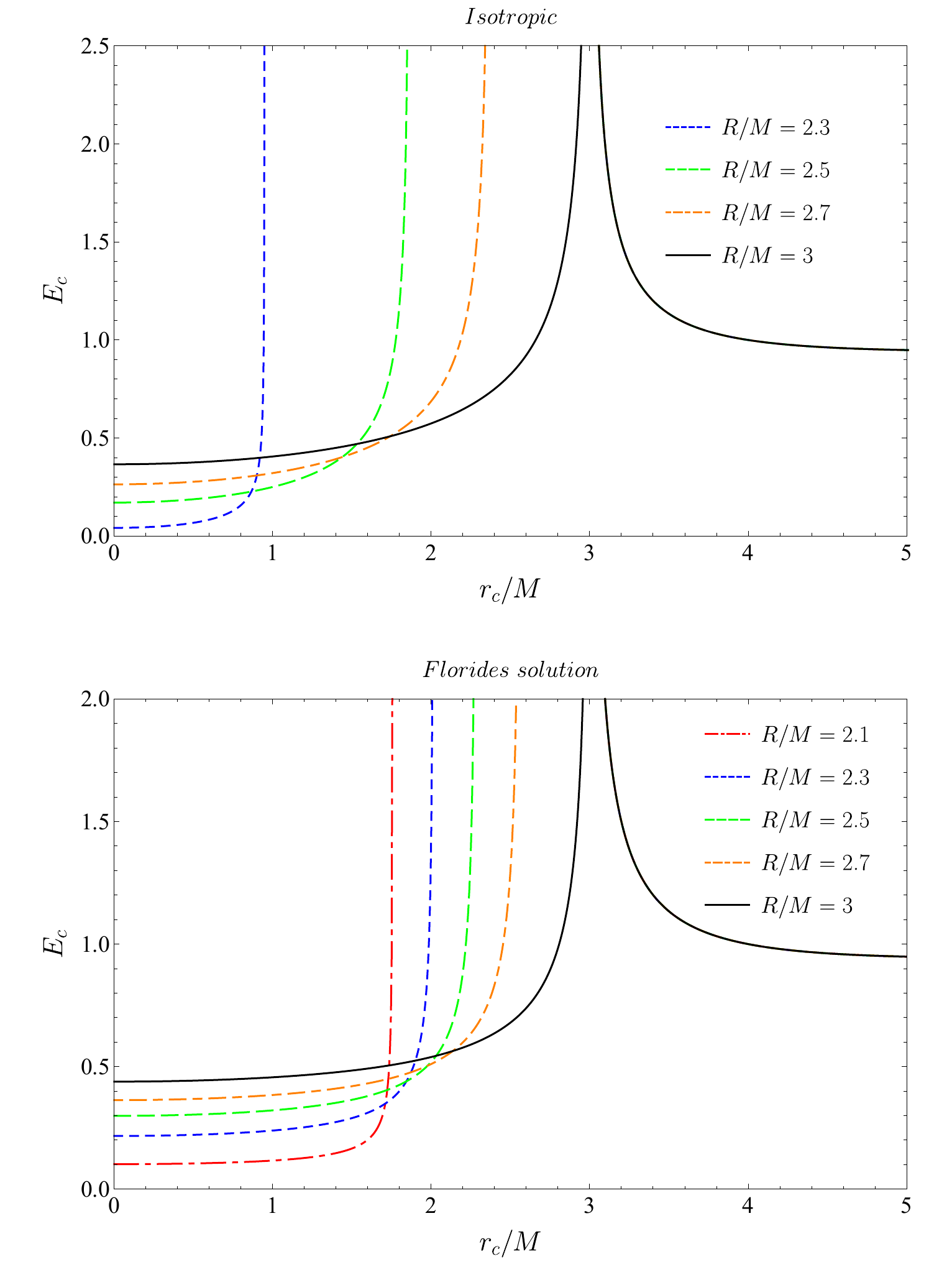}
    \caption{Specific energy for circular orbits around isotropic stars and for the Florides model for anisotropic objects. The curves are associated with different values of compactness $R/M$.}
    \label{fig:GrafE}
\end{figure}

\begin{figure}
    \centering
\includegraphics[width=1\linewidth]{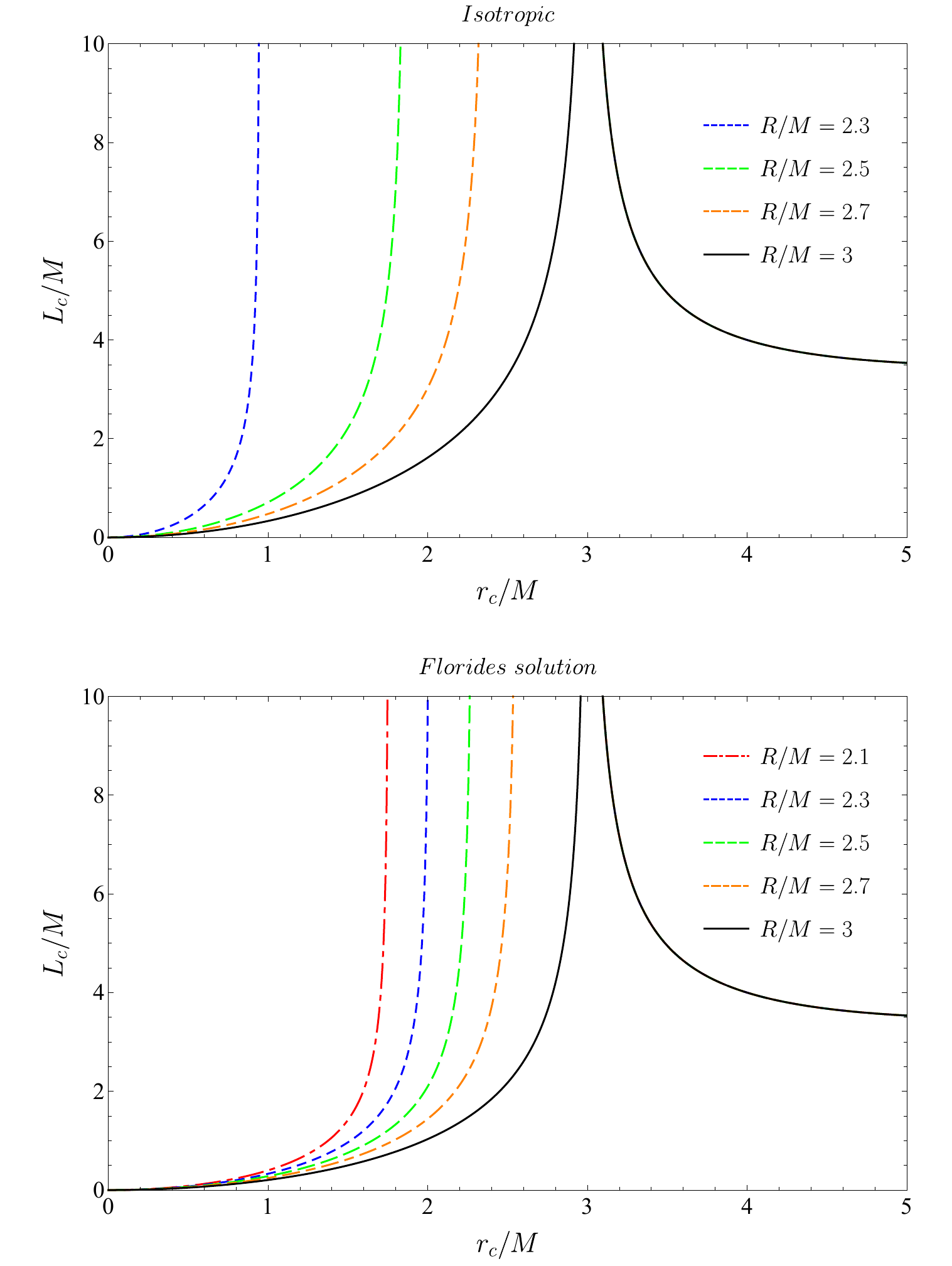}
    \caption{Specific angular momentum for circular orbits around isotropic stars and for the Florides model for anisotropic objects. The curves are associated with different values of compactness \( R/M \).}
    \label{fig:GrafL}
\end{figure}

Before entering into details of the tidal force's behavior, it is instructive to analyze the behavior of the specific energy and angular momentum for circular motion around the star. We can see in Figs. \ref{fig:GrafE} and \ref{fig:GrafL} how these quantities behave in the context of a circular orbit \textit{inside} and outside the ultracompact object. For different values of the star's radius, the values for the specific energy and angular momentum show the regions where it is possible to have circular time-like orbits. For the orbital motion inside the star, the values of $E_c$ and $L_c$ grow as $r_c$ approaches the stable null circular geodesic, where is formally diverge as predicted by Eqs.~\eqref{eq:e_l_timelike}. We notice that the second derivative of the effective potential is positive for these geodesics and, therefore, one can have ultrarelativistic \textit{stable} circular motion inside ultracompact objects.

\begin{figure}
    \centering
    \includegraphics[width=1\linewidth]{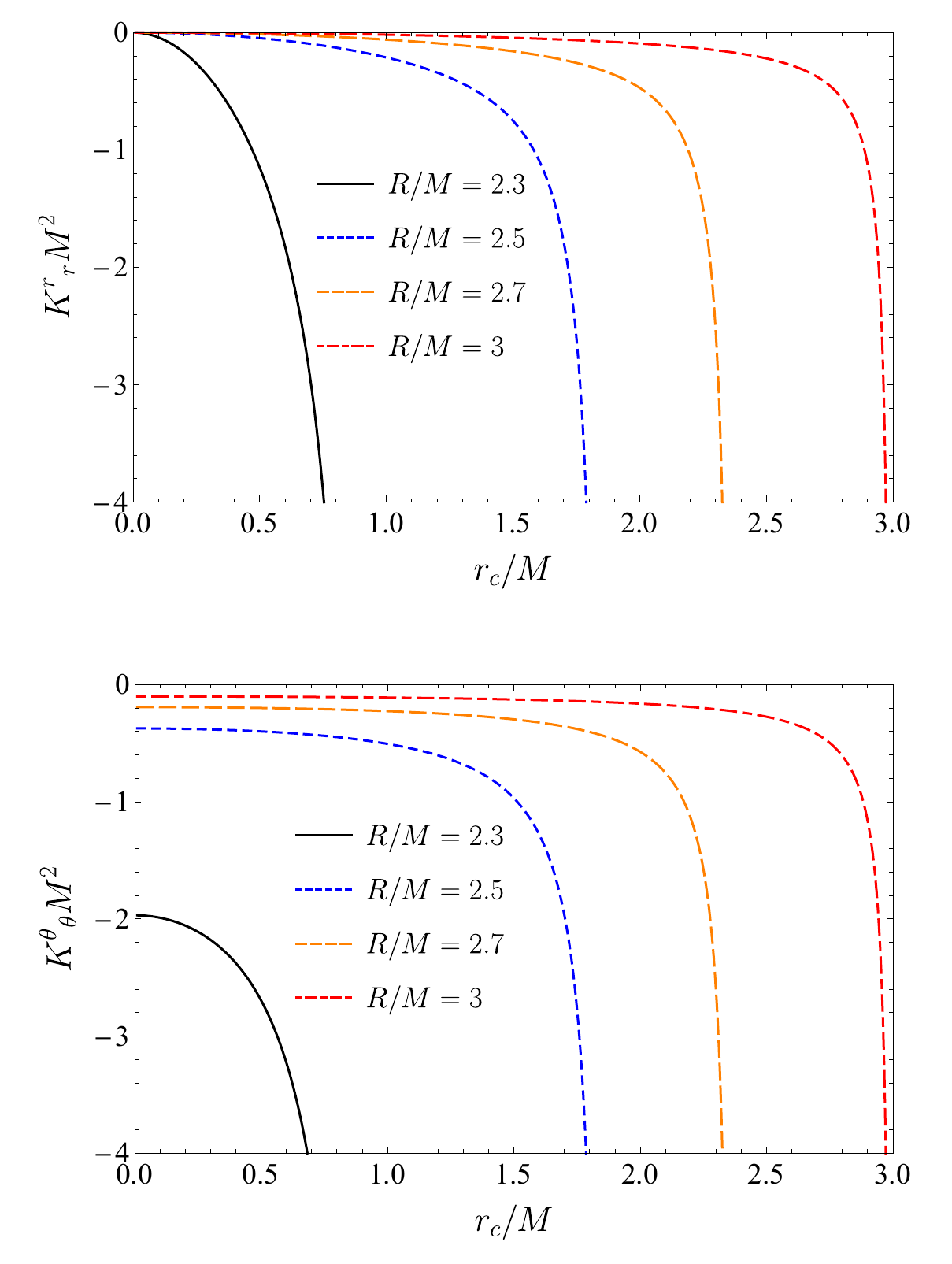}
    \caption{Radial and angular components of the tidal forces acting on a body in circular orbit inside an isotropic star. The curves are classified according to the star's compactness, $R/M$.}
    \label{fig:tfcisotropic}
\end{figure}

\begin{figure}
    \centering
    \includegraphics[width=1\linewidth]{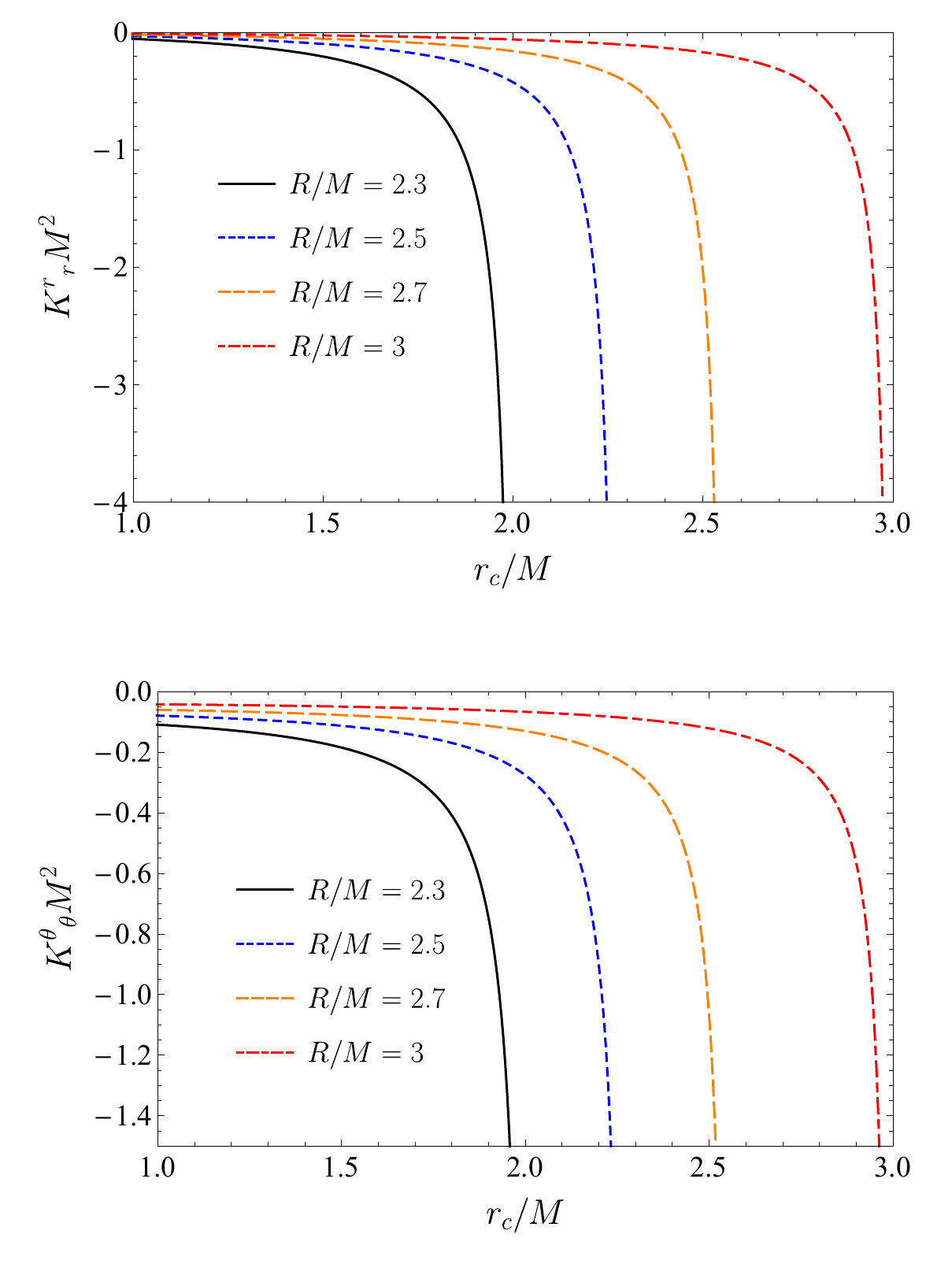}
    \caption{Radial and angular components of the tidal forces acting on a body in circular orbit inside an anisotropic star based on Florides' solution.  The curves are classified according to the star's compactness, $R/M$.}
    \label{fig:tfcflorides}
\end{figure}

Although ultrarelativistic motion within the star is formally stable, the particle limit of the geodesic should be view with care. We can verify the feasibility of having such orbital structures by looking into the tidal forces, as a sudden increase of them could prevent existent structure in such motion. In Figs. \ref{fig:tfcisotropic} and \ref{fig:tfcflorides}, we show the curvature component of the tidal forces -- as opposed to the centrifugal term -- for different values of compactness, for the isotropic and the Floride's solution, respectively. The behavior for the internal motion is evident: The divergences from the specific energy and angular momentum are clearly transferred to both radial and polar tidal forces. Since the azimuthal force only depends on inertial terms, it remains finite. Therefore, we conjecture that ultrarelativistic structures in circular orbits would not survive within the star.


The tidal force for orbital motions near the center of the star are important to analyze also the formation of structures that are accreted by it. For the radial tidal force, for both isotropic and Floride's cases, we see that the only contribution comes from the the inertial terms, with the curvature contributions vanishing identically as $r\to 0$. On the other hand, the polar contribution of the tidal force behaves differently for each stellar model. We find
\begin{align}
    \frac{d^2\xi^{\hat\theta}}{d\tau^2}&\approx-\frac{2M}{R^3(3\sqrt{1-2M/R}-1)}\xi^{\hat\theta} \hspace{.25cm}\mbox{ (isotropic),}\\
    \frac{d^2\xi^{\hat\theta}}{d\tau^2}&\approx -\frac{M}{R^3}\xi^{\hat\theta}\hspace{.25cm}\mbox{ (Florides),}
\end{align}
as $r\to0$. Clearly from the above, there is a divergent behavior at the center of the star for the isotropic case as $R$ goes to the Buchdahl limit. However, the polar tidal force on the Florides' case is perfectly regular at the center, just monotonically increasing with the compactness. The divergent behavior is more evidently seen from the expression in terms of the pressure, given by \eqref{eq:tfcamp2}, where we see a dependence with the radial pressure, which is known to diverge as we approach the Buchdahl limit at the center of the star. From these we conclude that the structure formation near the center of the star for extremely compact objects is model dependent.

\section{Conclusion}\label{sec:conclusion}

In this work, we have analyzed the relativistic tidal forces acting on circular geodesics around ultracompact objects, emphasizing their behavior near the null circular orbits, or light rings. By projecting the geodesic deviation equation onto the orthonormal frame of circular observers, we explicitly obtained the components of the tidal tensor and identified their divergence in the light-ring limit. This behavior is generic for spherically symmetric configurations, regardless of the details of the matter distribution.

To illustrate the physical consequences of these divergences, we examined two representative stellar models: the isotropic constant-density star, constrained by the Buchdahl bound, and the anisotropic Florides configuration, which allows higher compactness. In both cases, we found that the radial and polar components of the tidal acceleration diverge as the orbit approaches the inner and outer light rings, indicating that no stable bound structures could persist in these regions. This feature suggests that strong tidal gradients act as an effective mechanism preventing the long-term accumulation of matter or clumps near the photon orbits, thereby inhibiting potential nonlinear instabilities previously conjectured for ultracompact horizonless objects, such as the ones explored in Ref.~\cite{PhysRevD.90.044069}.

From a broader perspective, our results reinforce the view that light rings play a dual role in the dynamics of ultracompact objects: they are both signatures of extreme compactness and natural regulators of dynamical stability through the amplification of tidal stresses. Future work may explore the extension of this analysis to rotating configurations, dynamical perturbations, or alternative theories of gravity, where the interplay between light rings, tidal effects, and nonlinear instabilities could provide deeper insight into the nature and viability of black hole mimickers.

\begin{acknowledgements}
This work was supported by the Coordenação de Aperfeiçoamento de Pessoal de Nível Superior – Brasil (CAPES) – Finance Code 001, Fundação Amazônia de Amparo a Estudos e Pesquisa (FAPESPA), and Conselho Nacional de Desenvolvimento Científico e Tecnológico (CNPq).
\end{acknowledgements}

\bibliographystyle{ieeetr}

\bibliography{ref}

\begin{thebibliography}{10}

\bibitem{Testing_Cardoso_2019}
V.~Cardoso and P.~Pani, ``Testing the nature of dark compact objects: a status report,'' {\em Living Reviews in Relativity}, vol.~22, pp.~1--104, 2019.

\bibitem{2019}
K.~Akiyama~et al, ``First m87 event horizon telescope results. i. the shadow of the supermassive black hole,'' {\em The Astrophysical Journal Letters}, vol.~875, p.~L1, 2019.

\bibitem{PhysRevLett.116.061102}
B.~P. Abbott~et al, ``Observation of gravitational waves from a binary black hole merger,'' {\em Phys. Rev. Lett.}, vol.~116, p.~061102, 2016.

\bibitem{marks2025longtermstablenonlinearevolutions}
G.~A. Marks, S.~J. Staelens, T.~Evstafyeva, and U.~Sperhake, ``Long-term stable nonlinear evolutions of ultracompact black-hole mimickers,'' {\em Phys. Rev. Lett.}, vol.~135, p.~131402, 2025.

\bibitem{PhysRevLett.119.251102}
P.~V.~P. Cunha, E.~Berti, and C.~A.~R. Herdeiro, ``Light-ring stability for ultracompact objects,'' {\em Phys. Rev. Lett.}, vol.~119, p.~251102, 2017.

\bibitem{pani2009ergoregioninstabilityblackhole}
P.~Pani, V.~Cardoso, M.~Cadoni, and M.~Cavaglia, ``{Ergoregion instability of black hole mimickers},'' {\em PoS}, vol.~BHGRS, p.~027, 2008.

\bibitem{cunha2017light}
P.~V. Cunha, E.~Berti, and C.~A. Herdeiro, ``Light-ring stability for ultracompact objects,'' {\em Physical review letters}, vol.~119, p.~251102, 2017.

\bibitem{PhysRevD.105.064026}
J.~F.~M. Delgado, C.~A.~R. Herdeiro, and E.~Radu, ``Equatorial timelike circular orbits around generic ultracompact objects,'' {\em Phys. Rev. D}, vol.~105, p.~064026, 2022.

\bibitem{guo2024nonlinear}
G.~Guo, P.~Wang, and Y.-P. Zhang, ``Nonlinear stability of black holes with a stable light ring,'' {\em Phys. Rev. D}, vol.~112, p.~084023, 2025.

\bibitem{Rosa:2023qcv}
J.~L. Rosa, C.~F.~B. Macedo, and D.~Rubiera-Garcia, ``{Imaging compact boson stars with hot spots and thin accretion disks},'' {\em Phys. Rev. D}, vol.~108, p.~044021, 2023.

\bibitem{Rosa:2024bqv}
J.~L. Rosa, D.~S.~J. Cordeiro, C.~F.~B. Macedo, and F.~S.~N. Lobo, ``{Observational imprints of gravastars from accretion disks and hot spots},'' {\em Phys. Rev. D}, vol.~109, p.~084002, 2024.

\bibitem{cunha2025backreaction}
P.~V. Cunha, ``Backreaction of perturbations around a stable light ring,'' {\em Journal of Cosmology and Astroparticle Physics}, vol.~2025, p.~083, 2025.

\bibitem{Di_Filippo_2024}
F.~Di~Filippo, ``Nature of inner light rings,'' {\em Physical Review D}, vol.~110, p.~084026, 2024.

\bibitem{PhysRevD.90.044069}
V.~Cardoso, L.~C.~B. Crispino, C.~F.~B. Macedo, H.~Okawa, and P.~Pani, ``Light rings as observational evidence for event horizons: Long-lived modes, ergoregions and nonlinear instabilities of ultracompact objects,'' {\em Phys. Rev. D}, vol.~90, p.~044069, 2014.

\bibitem{Bizon:2011gg}
P.~Bizon and A.~Rostworowski, ``{On weakly turbulent instability of anti-de Sitter space},'' {\em Phys. Rev. Lett.}, vol.~107, p.~031102, 2011.

\bibitem{Lima_Junior_2022}
H.~C.~D. Lima~Junior, M.~M. Corrêa, C.~F.~B. Macedo, and L.~C.~B. Crispino, ``Tidal forces in dirty black hole spacetimes,'' {\em The European Physical Journal C}, vol.~82, p.~479, 2022.

\bibitem{chandrasekhar1969ellipsoidal}
S.~Chandrasekhar, {\em Ellipsoidal figures of equilibrium}.
\newblock New Haven and London: Yale University Press, 1969.

\bibitem{stockinger2024relativisticrocheproblemstars}
M.~Stockinger and M.~Shibata, ``Relativistic roche problem for stars in precessing orbits around a spinning black hole,'' {\em Phys. Rev. D}, vol.~110, p.~043038, 2024.

\bibitem{Wiggins_2000}
P.~Wiggins and D.~Lai, ``Tidal interaction between a fluid star and a kerr black hole in circular orbit,'' {\em The Astrophysical Journal}, vol.~532, p.~530, 2000.

\bibitem{1973ApJ...185...43F}
L.~G. {Fishbone}, ``{The Relativistic Roche Problem. I. Equilibrium Theory for a Body in Equatorial, Circular Orbit around a Kerr Black Hole},'' {\em The Astrophysical Journal}, vol.~185, pp.~43--68, 1973.

\bibitem{PhysRevD.99.104072}
G.~Raposo, P.~Pani, M.~Bezares, C.~Palenzuela, and V.~Cardoso, ``Anisotropic stars as ultracompact objects in general relativity,'' {\em Phys. Rev. D}, vol.~99, p.~104072, 2019.

\bibitem{chandrasekhar1998mathematical}
S.~Chandrasekhar, {\em The mathematical theory of black holes}.
\newblock Oxford university press, 1998.

\bibitem{carroll2019spacetime}
S.~M. Carroll, {\em Spacetime and geometry}.
\newblock Cambridge University Press, 2019.

\bibitem{hobson2006general}
M.~P. Hobson, G.~P. Efstathiou, and A.~N. Lasenby, {\em General relativity: an introduction for physicists}.
\newblock Cambridge university press, 2006.

\bibitem{d2023introducing}
R.~d'Inverno and J.~Vickers, {\em Introducing Einstein's relativity}.
\newblock : Oxford University Press, 2023.

\bibitem{PhysRev.55.364}
R.~C. Tolman, ``Static solutions of einstein's field equations for spheres of fluid,'' {\em Phys. Rev.}, vol.~55, pp.~364--373, 1939.

\bibitem{BeL}
R.~L. {Bowers} and E.~P.~T. {Liang}, ``{Anisotropic Spheres in General Relativity},'' {\em The Astrophysical Journal}, vol.~188, p.~657, 1974.

\bibitem{florides1974new}
P.~S. Florides, ``A new interior schwarzschild solution,'' {\em Proceedings of the Royal Society of London. A. Mathematical and Physical Sciences}, vol.~337, pp.~529--535, 1974.

\bibitem{PhysRev.116.1027}
H.~A. Buchdahl, ``General relativistic fluid spheres,'' {\em Phys. Rev.}, vol.~116, pp.~1027--1034, 1959.

\bibitem{Sharma_2021}
R.~Sharma, A.~Ghosh, S.~Bhattacharya, and S.~Das, ``Anisotropic generalization of buchdahl bound for specific stellar models,'' {\em The European Physical Journal C}, vol.~81, p.~527, 2021.

\bibitem{Fuchs90}
H.~Fuchs, ``Paralleltransport and geodesic deviation in static spherically symmetric space-times,'' {\em Astronomische Nachrichten}, vol.~311, pp.~219--222, 1990.

\bibitem{Madan:2022spd}
S.~Madan and P.~Bambhaniya, ``{Tidal force effects and bound orbits in null naked singularity spacetime},'' {\em Chin. Phys. C}, vol.~48, p.~115108, 2024.

\bibitem{Ghosh_2021}
R.~Ghosh and S.~Sarkar, ``Light rings of stationary spacetimes,'' {\em Physical Review D}, vol.~104, p.~044019, 2021.

\bibitem{Hong_2020}
S.-T. Hong, Y.-W. Kim, and Y.-J. Park, ``Tidal effects in schwarzschild black hole in holographic massive gravity,'' {\em Physics Letters B}, vol.~811, p.~135967, 2020.

\bibitem{thorne2000gravitation}
K.~S. Thorne, C.~W. Misner, and J.~A. Wheeler, {\em Gravitation}.
\newblock Freeman San Francisco, 2000.

\bibitem{mitsou2020tetrad}
E.~Mitsou and J.~Yoo, {\em Tetrad Formalism for Exact Cosmological Observables}.
\newblock Springer International Publishing, 2020.

\bibitem{zakharovTetradFormalismReference2006}
A.~F. Zakharov, V.~A. Zinchuk, and V.~N. Pervushin, ``Tetrad formalism and reference frames in general relativity,'' {\em Physics of Particles and Nuclei}, vol.~37, pp.~104--134, 2006.

\bibitem{Wulandari:2024gia}
N.~E.~S. Wulandari, B.~A. Subagyo, and M.~H. Rahmani, ``{Tetrad formalism in the solution of spherically symmetric spacetime in general relativity},'' {\em J. Phys. Conf. Ser.}, vol.~2780, p.~012029, 2024.

\bibitem{1983RSPSA.385..431M}
J.~A. {Marck}, ``{Solution to the Equations of Parallel Transport in Kerr Geometry; Tidal Tensor},'' {\em Proceedings of the Royal Society of London Series A}, vol.~385, pp.~431--438, 1983.

\bibitem{1990AN....311..271F}
H.~{Fuchs}, ``{Deviation of circular geodesics in static spherically symmetric space-times},'' {\em Astronomische Nachrichten}, vol.~311, pp.~271--276, 1990.

\bibitem{Philipp_2019}
D.~Philipp, D.~Puetzfeld, and C.~Lämmerzahl, {\em On the Applicability of the Geodesic Deviation Equation in General Relativity}, p.~419–451.
\newblock Springer International Publishing, 2019.

\bibitem{de_Andrade_2001}
V.~C. de~Andrade, L.~C.~T. Guillen, and J.~G. Pereira, ``Teleparallel spin connection,'' {\em Physical Review D}, vol.~64, p.~027502, 2001.

\end{thebibliography}

\end{document}